\documentclass[%
 reprint,
 amsmath,amssymb,
 aps,
floatfix,
]{revtex4-2}

\usepackage{amsmath}

\usepackage{xcolor}
\usepackage{graphicx}
\usepackage{dcolumn}
\usepackage{bm}
\usepackage{amsmath}
\usepackage{enumitem}
\usepackage{xcolor}
\usepackage{graphicx}
\usepackage{dcolumn}
\usepackage{bm}
\usepackage{float}

\begin{document}

\preprint{APS/123-QED}

\title{\textcolor{black}{Time-Domain and Spectral Decomposition of Dynamic Coupling with Instantaneous and Time-Lagged Causality: Methodological Evaluation and Application to Cardiovascular Variability}}



\author{Gorana Mijatovic}
 \altaffiliation{Faculty of Technical Sciences, University of Novi Sad, Serbia}
 \email{gorana86@uns.ac.rs}


 \author{Marta Iovino}
\altaffiliation{Department of Engineering, University of Palermo, Italy}
\email{marta.iovino@unipa.it}

 \author{Michal Javorka}
 \altaffiliation{Department of Physiology and Biomedical Center Martin, Jessenius Faculty of Medicine, Comenius University, Slovakia}
 \email{michal.javorka@uniba.sk}


 
\author{Luca Faes}
\altaffiliation{Department of Engineering, University of Palermo, Italy}
\email{luca.faes@unipa.it}


\begin{abstract}
The quantification of dynamic interactions in coupled physiological time series is a central question in biomedical signal processing. A critical issue is the presence of instantaneous dependencies, which are often neglected in conventional Granger causality (GC) analyses despite potentially reflecting physiologically meaningful dynamic effects. In this work, we systematically investigate time- and frequency-domain measures of causality and total coupling within the linear autoregressive (AR) framework, treating instantaneous dependencies according to three different formulations: (i) the classical strictly causal AR model, providing separate spectral measures of time-lagged GC and instantaneous interaction; (ii) the Geweke approach, providing spectral GC measures from extended AR models which assign instantaneous dependencies to the direction opposite to that under investigation; and (iii) a single extended AR model incorporating causally-oriented instantaneous dependencies, yielding spectral GC measures which account for both instantaneous and time-lagged causality. The three approaches are first evaluated in simulated Gaussian systems, for which the dynamic measures can be derived analytically and compared with the expected coupling structure. The single extended AR model emerges as the most consistent formulation when instantaneous dependencies have a physiologically meaningful causal direction and is subsequently applied to cardiovascular interactions between heart period (HP) and systolic (SAP) or diastolic arterial pressure (DAP) in healthy subjects during postural and mental stress. Accounting for instantaneous dependencies substantially affects the characterization of HP–SAP and HP–DAP interactions. In particular, their inclusion enables a better quantification of the baroreflex and cardiac run-off mechanisms, respectively, and provides a more refined characterization of their responses to stress and of low- and high-frequency cardiovascular dynamics. Overall, the proposed framework provides a consistent representation of causal interactions across time and frequency domains and highlights the importance of accounting for instantaneous dependencies in physiological coupling analysis.

\end{abstract}

\maketitle

\section {Introduction}

The assessment of interactions among dynamic processes represented by biomedical time series simultaneously measured from one or more organ systems is fundamental to understanding the underlying neurophysiological mechanisms. A prominent example is short-term cardiovascular regulation, where the closed-loop interplay between beat-to-beat variability of heart period (HP) and arterial pressure (AP, measured either as systolic arterial pressure (SAP) or diastolic arterial pressure (DAP)), reflects the combined action of neural autonomic reflexes and mechanical mechanisms operating over multiple temporal scales \cite{cohen2002short,javorka2017basic}.

Statistical methods for quantifying physiological interactions between two random processes observed as bivariate time series can be broadly classified according to the domain of analysis (time-domain versus spectral techniques \cite{malliani1991cardiovascular, faes2021information}), the type of interaction being evaluated (undirected or symmetric versus directed or causal measures \cite{schulz2013cardiovascular,porta2015wiener}), and the temporal lags at which interactions are assessed (zero-lag or static versus time-lagged or dynamic measures \cite{sparacino2025measuring,sparacino2025quantifying}).
Among the numerous approaches proposed in the literature, a prominent framework is that introduced by Geweke \cite{geweke1982measurement} based on the concept of Granger causality (GC) \cite{granger1969investigating}. Given two processes $X$ and $Y$, the framework defines logarithmic measures of causal (time-lagged) coupling along the two directions $X\rightarrow Y$ and $Y\rightarrow X$ which, together with a measure of instantaneous (zero-lag) coupling, provide a decomposition of the total interdependence between $X$ and $Y$. Owing to their computational simplicity coming from the fitting of bivariate autoregressive (AR) models, their straightforward interpretation in terms of improvement in prediction, and their close relationship with well-established information-theoretic measures \cite{chicharro2011spectral}, Geweke's GC measures have become a reference framework for assessing dynamic causal interactions in neurophysiological processes \cite{bressler2011wiener,porta2015wiener,pichot2024contribution}.

In contrast, instantaneous dependence has received comparatively less attention, mainly because it is not intrinsically associated with either causality or fully dynamic coupling. In fact, instantaneous effects have often been left unquantified, assumed to be absent, or disregarded in both theoretical and applied studies \cite{baccala2001partial, astolfi2007comparison, chicharro2011spectral, faes2013measuring}. However, interactions observed at zero lag may arise from causal relationships that cannot be resolved because of an insufficient sampling rate. This is particularly relevant in cardiovascular dynamics, which can be regarded as the output of a system paced by the heartbeat: in this setting, instantaneous dependencies reflect within-beat interactions that are causally oriented despite appearing as zero-lag effects in the sampled time series \cite{javorka2017basic,bara2026direct}. In the present study, we therefore consider an “extended” notion of causality \cite{schiatti2015extended} in which time-lagged and instantaneous causality are jointly quantified and their interpretation is informed by prior knowledge of the physiological system under investigation.

A further major issue addressed in this study concerns the assessment of causal relationships in the frequency domain when instantaneous effects are non-negligible and can be meaningfully assigned a causal direction. Spectral analysis of causal interactions is particularly valuable because it enables the characterization of directed influences within specific frequency bands of physiological relevance, such as the brain rhythms \cite{cao2022brain} and the low- and high-frequency oscillations observed in cardiovascular variability \cite{schulz2013cardiovascular}. Accordingly, frequency-domain formulations of causal interactions have been developed using a variety of approaches \cite{astolfi2007comparison, porta2015wiener}.
Many of these approaches rely on the spectral representation of \textit{strictly causal} AR models, i.e., models that do not explicitly account for instantaneous effects and therefore neglect their contribution to the coupling structure.
However, considering that the presence of zero-lag interactions has been shown to substantially affect spectral measures of coupling and causality \cite{faes2013framework,baccala2021frequency}, several modifications of existing approaches have been introduced to quantify causality in the spectral domain while accounting for instantaneous effects. These include formulations based on \textit{extended} AR models incorporating zero-lag effects into the model coefficients \cite{geweke1982measurement,faes2013framework}, approaches based on the comparison between full and “disconnected” time series models \cite{cohen2020general,nuzzi2021extending}, and methods providing an independent spectral characterization of instantaneous effects \cite{baccala2021frequency,pernice2022spectral}.

In the present work, we investigate measures of causality and total coupling in the frequency domain under different treatments of instantaneous effects. Specifically, we consider three approaches: (i) the spectral expansion of strictly causal AR models, which provides logarithmic, frequency-specific measures of GC and instantaneous interaction whose sum yields a spectral measure of total dependence, nonlinearly related to the squared magnitude coherence \cite{ding200617,pernice2022spectral}; (ii) the spectral formulation proposed by Geweke \cite{geweke1982measurement,bressler2011wiener}, in which two different extended AR models are used to quantify spectral GC after suppressing instantaneous effects in the direction opposite to that under investigation; and (iii) an approach based on a single extended AR model that explicitly incorporates causally meaningful zero-lag influences, which is represented in the frequency domain yielding two spectral GC measures (one exclusively time-lagged and one augmented with instantaneous causality) whose sum provides a new measure of total dependence.

The three approaches are first evaluated in simulated Gaussian systems, for which the exact theoretical profiles of the spectral measures and their corresponding time-domain counterparts can be derived analytically. This allows us to assess their ability to correctly identify causal and total coupling and to compare their frequency-specific representations with the expected theoretical results. Subsequently, the measures identified as the most appropriate for representing causal and total coupling in the time and frequency domains are applied to a physiological setting in which instantaneous causality is expected to play a relevant role. Specifically, we consider cardiovascular interactions obtained from AR modeling of HP$–$SAP and HP$–$DAP bivariate interactions in a benchmark dataset of physiological time series recorded from healthy subjects during postural and mental stress \cite{saputo2026partial}.

\section {Methods}
\label{sec_methods}

This work considers a discrete-time bivariate dynamic process expressed as
$\mathbf{S}=\{X,Y\}$, where $X=\{X_n\}_{n \in \mathbb{Z}}$ and $Y=\{Y_n\}_{n \in \mathbb{Z}}$ are zero-mean stationary stochastic processes.
In this section we present three different approaches for the computation of measures of causal (i.e., directed from $X$ to $Y$ or vice versa) and total (i.e., symmetric) dynamic coupling between the two analyzed processes, expressed in both time and frequency domains and obtained within the frame of Granger causality (GC)   \cite{granger1969investigating}.
Specifically, we consider first the computation based on the classical  strictly causal autoregressive (AR) representation of the bivariate process, \cite{faes2012measuring} implemented by treating separately time-lagged (causal) and zero-lag (instantaneous) interactions between $X$ and $Y$ \cite{pernice2022spectral} (Sect. \ref{subsec_A}). Then, we describe the formulation proposed by Geweke \cite{geweke1982measurement} to obtain frequency-domain GC accounting exclusively for time-lagged effects and discounting the instantaneous interaction (Sect. \ref{subsec_B}). Finally, we present an approach to infer time- and frequency-domain GC when the instantaneous interaction is considered causally meaningful and aggregated to time-lagged effects (Sect. \ref{subsec_C}).

\subsection{Time- and frequency-domain interaction measures from single strictly causal AR model}
\label{subsec_A}

The most common approach to quantify GC in the time domain is based on modeling the joint dynamics of $\mathbf{S}$ by means of a bivariate autoregressive (AR) model of order $p$,
\begin{subequations} \label{eq_AR}  
\begin{align}
X_n &= \sum_{k=1}^{p} a_{11,k} X_{n-k} + \sum_{k=1}^{p} a_{12,k} Y_{n-k} + U_{1,n} \label{eq_VAR_a}, \\
Y_n &= \sum_{k=1}^{p} a_{21,k} X_{n-k} + \sum_{k=1}^{p} a_{22,k} Y_{n-k} + U_{2,n} \label{eq_VAR_b},
\end{align}
\end{subequations}
\noindent where the model coefficients $a_{ij,k}$ describe self-dependencies ($i=j$), capturing how the past $k$ values of an individual process influence its current state at time $n$, as well as cross-dependencies ($i \neq j$), capturing the influence of past values of one process on the current state of the other, for $k = 1,\dots,p$. The innovation processes $U_{1}$ and $U_{2}$ account for stochastic variations of $\mathbf{S}$ not explained by its past values; they have zero mean and are temporally uncorrelated, so that their joint representation $\mathbf{U} = [U_{1}, U_{2}]^\mathrm{T}$ is fully provided by the covariance matrix $\boldsymbol{\Sigma} = \mathbb{E}[\mathbf{U}_n \mathbf{U}_n^\mathrm{T}] = \begin{bmatrix} \sigma_1^2 & \sigma_{12} \\ \sigma_{12} & \sigma_2^2 \end{bmatrix}$, 
where $\sigma_1^2 = \mathbb{E}[U_{1,n}^2]$, $\sigma_2^2 = \mathbb{E}[U_{2,n}^2]$, $\sigma_{12} = \sigma_{21} = \mathbb{E}[U_{1,n} U_{2,n}]$; $\mathbb{E}[\cdot]$ denotes the expectation operator.

\begin{figure} [t!]
\centering
\includegraphics[scale=1]{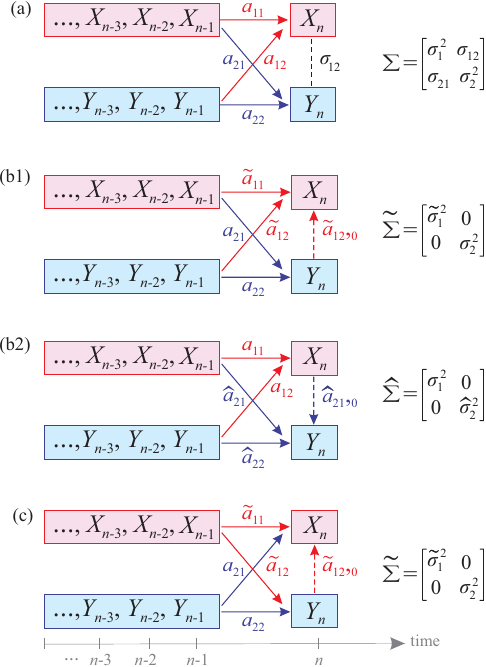}
\caption{
Conceptual representation of the three AR model formulations considered in this work.
(a) Strictly causal bivariate AR model, where only time-lagged effects are modeled and instantaneous interactions are implicitly represented by the off-diagonal elements of the innovation covariance matrix $\mathbf{\Sigma}$.
(b) Geweke's orthogonalized extended models, in which the instantaneous interaction is explicitly modeled through a zero-lag coefficient, yielding an innovation covariance matrix with uncorrelated innovations (diagonal innovation covariance $\mathbf{\tilde{\Sigma}}$); two orthogonalizations are possible, in which the instantaneous interaction is assigned the direction $Y\to X$ (b1) or the direction $X\to Y$ (b2).
(c) Single orthogonalized model, where the instantaneous interaction (known \textit{a priori}) is assigned to only one direction (here, $Y\to X$) and the innovation covariance is diagonal; this model is identical to model (b1), but used to compute GC in both directions.
}
\label{fig:concept}
\end{figure}

The bivariate AR model (\ref{eq_AR}) is often denoted as \textit{strictly causal}, meaning that all the interactions described in terms of the model coefficients are time-lagged (i.e., $k > 0$ in both (\ref{eq_VAR_a}) and (\ref{eq_VAR_b})). As a consequence, instantaneous interactions between $X$ and $Y$ (i.e., interactions occurring at lag $k = 0$), when present, remain as interactions between the model residuals and are quantified by the off-diagonal terms of the innovation covariance matrix, $\sigma_{12} = \sigma_{21}$, as illustrated in Fig. \ref{fig:concept}(a).


To derive measures of causal interaction, the treatment is completed considering the univariate AR representation for each process,
\begin{subequations} \label{eq_AR_restricted}
\begin{align}
X_n = \sum_{k=1}^{\infty} b_{1,k} X_{n-k} + W_{1,n}, \\
Y_n = \sum_{k=1}^{\infty} b_{2,k} Y_{n-k} + W_{2,n}, 
\end{align}
\end{subequations}
\noindent describing the intrinsic (self-driven) dynamics of $X$ and $Y$, whereby the current value of a process depends exclusively on its own past through the coefficients $b_{1,k}$ and $b_{2,k}$ for $k > 0$; the residual terms $W_{1,n}$ and $W_{2,n}$ are zero-mean white noise innovation processes with variances $\lambda_1^2 = \mathbb{E}[W_{1,n}^2]$ and $\lambda_2^2 = \mathbb{E}[W_{2,n}^2]$. Note that the order of the reduced AR models (\ref{eq_AR_restricted}a, \ref{eq_AR_restricted}b) derived from the full model (\ref{eq_AR}) typically tends to infinite because a finite-order bivariate AR model cannot be expressed in closed form using univariate AR models
\cite{barnett2015granger}.

Given the model representations in (\ref{eq_AR}a, \ref{eq_AR}b) and (\ref{eq_AR_restricted}a, \ref{eq_AR_restricted}b), the time-lagged and instantaneous interaction between the two processes can be quantified separately using the logarithmic measures proposed by Geweke \cite{geweke1982measurement}. In particular, directed time-lagged interactions are defined through the Granger causality (GC) framework, capturing the influence of the driving process $X$ on the target $Y$ and, conversely, of the driver $Y$ on the target $X$, respectively, as
\begin{subequations} \label{GCtime}
\begin{align}
F_{X \to Y} &\triangleq \ln \frac{\lambda_2^2}{\sigma_2^2}, \label{GCtimea} \quad \\
F_{Y \to X} &\triangleq \ln \frac{\lambda_1^2}{\sigma_1^2},
\label{GCtimeb}
\end{align}
\end{subequations}
\noindent where each measure represents the reduction in the prediction error variance of the target process achieved by incorporating the past of the driving process in the prediction model. The instantaneous coupling between the processes is instead quantified as
\begin{equation} \label{ICtime}
F_{X \cdot Y} \triangleq \ln \frac{\sigma_1^2 \sigma_2^2}{|\mathbf{\Sigma}|},
\end{equation}
\noindent where $|\boldsymbol{\Sigma}|$ denotes the determinant of the innovation covariance matrix. The sum of these components yields the total interdependence between $X$ and $Y$:
\begin{equation} 
F_{X;Y} \triangleq F_{X \to Y} + F_{Y \to X} + F_{X \cdot Y}
= \ln \frac{\lambda_1^2 \lambda_2^2}{|\mathbf{\Sigma}|}.
\label{eq:gc_total}
\end{equation}

All of these measures are non-negative and provide complementary information on the interactions between the processes: the directed measures, $F_{X \to Y}$ and $F_{Y \to X}$, quantify the causal influence between the processes according to the GC principle \cite{granger1969investigating}, whereas $F_{X \cdot Y}$ is undirected (i.e., $F_{X \cdot Y}=F_{Y \cdot X}$) and captures instantaneous (zero-lag) correlations between the processes. The total interdependence, $F_{X;Y}=F_{Y;X}$, aggregates these contributions, vanishing when the processes are independent (i.e., when the AR coefficients representing cross-dependencies are all null and the innovation covariance matrix is diagonal) and increasing with both directed and instantaneous interactions. Crucially, for jointly Gaussian processes, these measures own a straightforward information-theoretic interpretation, with the total interdependence corresponding to twice the mutual information rate, the directed GC measures to twice the transfer entropy, and the instantaneous coupling to twice the conditional mutual information between the present states given the past \cite{barnett2009granger, chicharro2011spectral, faes2022new}.

\bigskip

The bivariate AR model in (\ref{eq_AR}) admits a frequency-domain representation, allowing the interaction measures to be decomposed across frequencies. Following the framework proposed in \cite{faes2022new, pernice2022spectral}, the spectral representation is obtained by expressing the AR model through its transfer function and relating it to the power spectral density (PSD) matrix of the observed processes. Specifically, applying the $Z$-transform to (\ref{eq_VAR_a},\ref{eq_VAR_b}), with $z=e^{j\Omega}$, and rearranging terms, the model can be written as $\mathbf{A}(z)\mathbf{S}(z)=\mathbf{U}(z)$, or equivalently as
\begin{equation}
\label{eq:matrix_freq}
\begin{bmatrix}
A_{11}(z) & A_{12}(z)\\
A_{21}(z) & A_{22}(z)
\end{bmatrix}
\begin{bmatrix}
X(z)\\
Y(z)
\end{bmatrix}
=
\begin{bmatrix}
U_1(z)\\
U_2(z)
\end{bmatrix},
\end{equation}

\noindent where 
$A_{11}(z)=1-\sum_{k=1}^{p}a_{11,k}z^{-k}
$, $A_{12}(z)=-\sum_{k=1}^{p}a_{12,k}z^{-k}$, $A_{21}(z)=-\sum_{k=1}^{p}a_{21,k}z^{-k}$, and $A_{22}(z)=1-\sum_{k=1}^{p}a_{22,k}z^{-k}$;  $\mathbf{S}(z)=[X(z)\;Y(z)]^\top$, and $\mathbf{U}(z)=[U_1(z)\;U_2(z)]^\top$. Further, introducing the transfer matrix
\begin{equation}
\mathbf{H}(z)=\mathbf{A}^{-1}(z),
\end{equation}

\noindent leads to the vector moving-average representation of the model, defined as $ \mathbf{S}(z)=\mathbf{H}(z)\mathbf{U}(z)$, or equivalently as

\begin{equation}
\begin{bmatrix}
X(z)\\
Y(z)
\end{bmatrix}
=
\begin{bmatrix}
H_{11}(z) & H_{12}(z)\\
H_{21}(z) & H_{22}(z)
\end{bmatrix}
\begin{bmatrix}
U_1(z)\\
U_2(z)
\end{bmatrix}.
\end{equation}

\noindent Evaluating the transfer matrix on the unit circle of the complex plane, i.e., setting $z=e^{j\Omega}$, where $\Omega=2\pi f/f_s$ is the normalized angular frequency (with $f$ frequency and $f_s$ sampling frequency), the PSD matrix of the bivariate process is obtained through the important spectral factorization formula \cite{kay1988modern}:
\begin{equation}
\mathbf{P}_\mathrm{\textbf{S}}(\Omega)
=
\mathbf{H}(\Omega)
\boldsymbol{\Sigma}
\mathbf{H}^{*}(\Omega),
\label{eq_PSD}
\end{equation}
where $(\cdot)^*$ denotes the Hermitian transpose.
The diagonal elements of $\mathbf{P}_\mathrm{\textbf{S}}(\Omega)$ define the power spectral densities of the two processes as
\begin{align}
P_X(\Omega)
&=
\sigma_1^2|H_{11}(\Omega)|^2
+
\sigma_{12}H_{11}(\Omega)H_{12}^{*}(\Omega)
\nonumber\\
&
+
\sigma_{21}H_{12}(\Omega)H_{11}^{*}(\Omega)
+
\sigma_2^2|H_{12}(\Omega)|^2,
\label{eq_PX}
\\
P_Y(\Omega)
&=
\sigma_1^2|H_{21}(\Omega)|^2
+
\sigma_{12}H_{21}(\Omega)H_{22}^{*}(\Omega)
\nonumber\\
&
+
\sigma_{21}H_{22}(\Omega)H_{21}^{*}(\Omega)
+
\sigma_2^2|H_{22}(\Omega)|^2,
\label{eq_PY}
\end{align}

\noindent while the off-diagonal elements define the cross-spectrum
$P_{XY}(\Omega)$ and its complex conjugate
$P_{YX}(\Omega)=P_{XY}^{*}(\Omega)$.

\bigskip

The frequency-domain counterparts of the interaction measures are then obtained from the spectral density matrix. In particular, spectral GC measures from $X$ to $Y$ and from $Y$ to $X$ can be defined, respectively, as \cite{faes2022new, pernice2022spectral}
\begin{equation}
f_{X\rightarrow Y}(\Omega)
\triangleq
\ln
\frac{P_Y(\Omega)}
{\sigma_2^2|H_{22}(\Omega)|^2},
\label{eq_fxy}
\end{equation}

\begin{equation}
f_{Y\rightarrow X}(\Omega)
\triangleq
\ln
\frac{P_X(\Omega)}
{\sigma_1^2|H_{11}(\Omega)|^2}.
\label{eq_fyx}
\end{equation}

\noindent Moreover, the spectral representation of the instantaneous interaction between $X$ and $Y$ is defined as \cite{faes2022new, pernice2022spectral}
\begin{equation}
f_{X\cdot Y}(\Omega)
\triangleq
\ln
\frac{\sigma_1^2\sigma_2^2
|H_{11}(\Omega)|^2
|H_{22}(\Omega)|^2}
{|\mathbf{P}_\mathbf{\textbf{S}}(\Omega)|},
\label{eq_finst}
\end{equation}
\noindent where $|\mathbf{P}_\mathbf{\textbf{S}}(\Omega)|$ denotes the determinant of the spectral density matrix. Accordingly, the total interaction is defined as the sum of the three spectral components,
\begin{equation}
f_{X;Y}(\Omega)\triangleq f_{X\rightarrow Y}(\Omega) + f_{Y\rightarrow X}(\Omega) + f_{X\cdot Y}(\Omega),
\label{dec_freq_enforced}
\end{equation}
yielding
\begin{equation} \label{f_XY_freq}
f_{X;Y}(\Omega)
=
\ln
\frac{P_X(\Omega)P_Y(\Omega)}
{|\mathbf{P}_\mathbf{\textbf{S}}(\Omega)|},
\end{equation}
which, using the definition of squared coherence, can be equivalently expressed as
\begin{equation}
f_{X;Y}(\Omega)
=
-\ln\!\left(1-\Gamma_{X;Y}^{2}(\Omega)\right),
\label{eq:coh}
\end{equation}
where
\begin{equation}
\Gamma_{X;Y}^{2}(\Omega)
\triangleq
\frac{|P_{XY}(\Omega)|^2}
{P_X(\Omega)P_Y(\Omega)}
\label{eq_coherence}
\end{equation}
is the well-known magnitude squared coherence between the two processes \cite{kay1988modern}.

Thus, the spectral interaction measures defined in (\ref{eq_fxy}-\ref{eq_finst}) preserve the additive decomposition established by Geweke in the time domain (Eq. (\ref{eq:gc_total})). Another desirable property is that the integration of each of the spectral measures
$f_{X\rightarrow Y}(\Omega)$, $f_{Y\rightarrow X}(\Omega)$,
$f_{X\cdot Y}(\Omega)$, and $f_{X;Y}(\Omega)$ over the entire frequency range recovers the corresponding time-domain measures
$F_{X\rightarrow Y}$, $F_{Y\rightarrow X}$,
$F_{X\cdot Y}$, and $F_{X;Y}$ \cite{chicharro2011spectral,faes2022new}; this makes a link between the time- and frequency-domain measures defined in (\ref{GCtime}-\ref{ICtime}) and in (\ref{eq_fxy}-\ref{eq_finst}).
However, the spectral measures can take negative values at some frequencies, which impairs their interpretability. This issue, which is documented in the simulation section, holds in the presence of non-negligible instantaneous interaction between the processes ($\sigma_{12}>0$) for the two GC measures defined in (\ref{eq_fxy}) and (\ref{eq_fyx}); in the absence of instantaneous dependencies the two GCs are non-negative at each frequency \cite{faes2022new}. On the other hand, the negativity issue holds even for independent innovations ($\sigma_{12}=0$) for the spectral measure (\ref{eq_finst}), reflecting the fact that this measure does not have a clear interpretation on its own but was rather defined to enforce the spectral decomposition (\ref{dec_freq_enforced}). An alternative approach to the spectral representation of causal interactions, which overcomes the negativity issue, is presented in the next subsection.


\subsection{Frequency-domain Granger causality measures from different extended AR models}
\label{subsec_B}

To implement GC measures in the frequency domain, Geweke \cite{geweke1982measurement} proposed two orthogonalized representations of the bivariate AR model (\ref{eq_AR}) which include a term accounting for the so-called \textit{instantaneous causality}. Specifically, with this approach an \textit{extended} AR model incorporating  instantaneous causality along the direction $Y_n \rightarrow X_n$ is used to quantify the spectral GC from $X$ to $Y$ (Fig. 1(b1)), and an extended model incorporating  instantaneous causality along the direction $X_n \rightarrow Y_n$ is used to quantify the spectral GC from $Y$ to $X$ (Fig. 1(b2)). As a result, the instantaneous interaction is shifted from the innovation covariance matrix (which indeed becomes diagonal) to the model coefficient quantifying instantaneous causality (i.e., zero-lag interaction with an assigned direction).

In detail, to quantify the spectral GC from $X$ to $Y$, a modified bivariate AR model is introduced in which the description of $X$ is augmented by introducing in (\ref{eq_VAR_a}) a term accounting for the instantaneous causality  $Y_n \rightarrow X_n$, while the description of $Y$ remains unchanged as in (\ref{eq_VAR_b}). This
 extended model is defined by applying the Cholesky decomposition to (\ref{eq_AR}), which yields:
 \begin{subequations} \label{eq_AR_inst}
\begin{align}
X_n &= \sum_{k=1}^{p} \tilde{a}_{11,k} X_{n-k} + \sum_{k=0}^{p} \tilde{a}_{12,k} Y_{n-k} + \tilde{U}_{1,n}, \label{eq_AR_inst_a} \\
Y_n &= \sum_{k=1}^{p} a_{21,k} X_{n-k} + \sum_{k=1}^{p} a_{22,k} Y_{n-k} + U_{2,n}, \label{eq_AR_inst_b}
\end{align}
\end{subequations}
\noindent where the modified coefficients are 
\(\tilde a_{11,k} = a_{11,k} - \sigma_{12}/\sigma_2^2 \cdot a_{21,k}\) and 
\(\tilde a_{12,k} = a_{12,k} - \sigma_{12}/\sigma_2^2 \cdot a_{22,k}\) for $k>0$ and where the new coefficient $\tilde a_{12,0}=\sigma_{12}/\sigma_2^2$ explicitly captures the zero-lag effect of \(Y_n\) on \(X_n\). The innovation process becomes
\(\tilde U_{1,n} = U_{1,n} - \sigma_{12}/\sigma_2^2 \cdot U_{2,n}\), 
which implies that the innovations are uncorrelated, i.e., 
\(\mathbb{E}[\tilde U_{1,n} U_{2,n}] = 0\). This produces a new innovation covariance matrix 
\(\boldsymbol{\tilde\Sigma} = \mathbb{E}[\tilde{\mathbf U}_n \tilde{\mathbf U}_n^\mathrm{T}] = 
\begin{bmatrix} 
\tilde \sigma_1^2 & 0 \\ 
0 & \sigma_2^2 
\end{bmatrix}\), 
where 
\begin{equation} \label{tildasigma1}
    \tilde \sigma_1^2 = \mathbb{E}[\tilde U_{1,n}^2] = \sigma_1^2 - \frac{\sigma_{12}^2}{\sigma_2^2},
\end{equation}
\noindent while \(\sigma_2^2\) remains unchanged (Fig. 1(b1)).

The orthogonalized representation (\ref{eq_AR_inst}) provides the basis for the frequency-domain characterization of the Granger causality from $X$ to $Y$. Specifically, 
applying the $Z$-transform to (\ref{eq_AR_inst_a}, \ref{eq_AR_inst_b}) yields $\mathbf{\tilde A}(z) \mathbf{S}(z)=  \mathbf{\tilde U}(z)$, or equivalently
\begin{equation} \label{eq: matrix_form}
\begin{bmatrix}
\tilde{A}_{11}(z) & \tilde{A}_{12}(z) \\
A_{21}(z) & A_{22}(z)
\end{bmatrix}
\begin{bmatrix}
X(z) \\
Y(z)
\end{bmatrix}
=
\begin{bmatrix}
\tilde{U}_1(z) \\
U_2(z)
\end{bmatrix},
\end{equation}
where $\tilde{A}_{11}(z)=1 - \sum_{k=1}^{p} \tilde{a}_{11,k} z^{-k}$, $\tilde{A}_{12}(z)=- \sum_{k=0}^{p} \tilde{a}_{12,k} z^{-k}$, $A_{21}(z)=-\sum_{k=1}^{p}a_{21,k}z^{-k}$, and $A_{22}(z)=1-\sum_{k=1}^{p}a_{22,k}z^{-k}$; $\mathbf{S}(z)=[X(z)\;Y(z)]^\top$, and $\mathbf{\tilde U}(z)=[\tilde{U}_1(z)\;U_2(z)]^\top$.
\noindent Accordingly, the transfer matrix is defined as
\begin{equation}
\tilde{\mathbf{H}}(z) = \tilde{\mathbf{A}}^{-1}(z),
\end{equation}
\noindent leading to the moving-average representation
$\mathbf{S}(z)=\mathbf{\tilde H}(z)\mathbf{\tilde U}(z)$. Given this representation, evaluating the transfer matrix on the unit circle
($z=e^{j\Omega}$) yields the alternative formulation of the PSD matrix:
\begin{equation} \label{PSD_S}
\mathbf{P_S}(\Omega) = \tilde{\mathbf{H}}(\Omega)\,\tilde{\boldsymbol{\Sigma}}\,\tilde{\mathbf{H}}^{*}(\Omega),
\end{equation}
\noindent from which the PSD of the process $Y$ is obtained as the lower diagonal element, and can be written as
\begin{equation} \label{P_Y}
P_Y(\Omega) = |H_{21}(\Omega)|^2 \tilde{\sigma}_1^2 + |\tilde H_{22}(\Omega)|^2 \sigma_2^2.
\end{equation}
\noindent Then, the spectral GC measure from $X$ to $Y$ can be defined similarly to (\ref{eq_fxy}) as \cite{geweke1982measurement}
\begin{equation}\label{fxy_ext}
\tilde{f}_{X\rightarrow Y}(\Omega)
\triangleq
\ln
\frac{P_Y(\Omega)}
{|\tilde H_{22}(\Omega)|^2\sigma_2^2}
=
\ln
\frac{P_Y(\Omega)}
{P_Y(\Omega)-|H_{21}(\Omega)|^2\tilde\sigma_1^2}.
\end{equation}

Afterwards, to quantify the spectral GC along the opposite causal direction $Y \rightarrow X$, the same procedure is applied starting from the alternative orthogonalized representation, in which (\ref{eq_VAR_b}) is modified while (\ref{eq_VAR_a}) remains unchanged. This corresponds to using an extended model analogous to (\ref{eq_AR_inst}) where the roles of $X$ and $Y$ are swapped (Fig. 1(b2)), which intuitively leads to the spectral factorization 
\begin{equation}
\mathbf{P_S}(\Omega) = \hat{\mathbf{H}}(\Omega)\,\hat{\boldsymbol{\Sigma}}\,\hat{\mathbf{H}}^{*}(\Omega),
\end{equation}
from which the PSD of the target and the spectral GC from $Y$ to $X$ are obtained as
\begin{equation} \label{P_X}
P_X(\Omega) = |\hat{H}_{11}(\Omega)|^2 \sigma_1^2 + |H_{12}(\Omega)|^2 \hat{\sigma}_2^2,
\end{equation}
\begin{equation} \label{fyx_ext}
\hat{f}_{Y\rightarrow X}(\Omega)
\triangleq
\ln
\frac{P_X(\Omega)}
{|\hat{H}_{11}(\Omega)|^2\sigma_1^2}
=
\ln
\frac{P_X(\Omega)}
{P_X(\Omega)-|{H}_{12}(\Omega)|^2\hat{\sigma}_2^2},
\end{equation}
\noindent where the modified partial innovation variance is $\hat{\sigma}_2^2 = \sigma_2^2 - \sigma_{12}^2 / \sigma_1^2 $.

The spectral GC measures $\tilde{f}_{X\rightarrow Y}(\Omega)$ and $\hat{f}_{Y\rightarrow X}(\Omega)$ defined in (\ref{fxy_ext}) and (\ref{fyx_ext}) satisfy the spectral integration property such that their integration over the whole frequency range recovers the time-domain GC measures $F_{X\rightarrow Y}$ and $F_{Y\rightarrow X}$ \cite{geweke1982measurement}. Moreover, they are always non-negative at each frequency by construction, thanks to the orthogonalization which makes the individual PSDs composed of two non-negative terms (see (\ref{P_Y}) and (\ref{P_X})).
However, being derived from two different models the spectral measures do not result from a joint representation of the dynamics of the bivariate process $\mathbf{S}=\{X,Y\}$; for this reason a relation similar to (\ref{f_XY_freq}) does not hold for these measures, and their sum cannot be interpreted as a measure of total spectral coupling. A way to overcome this limitation is presented in the next subsection.

\subsection{Time- and frequency-domain interaction measures from single extended AR model}
\label{subsec_C}

Here, we propose a formulation of the measures of causal and total dynamic interaction in bivariate processes, expressed both in the time and frequency domains, which joins the advantages of the approaches presented in Sect. II.A and II.B. Specifically, under the hypothesis that a causal interpretation can be given to the instantaneous interaction between $X$ and $Y$, and that the direction of this zero-lag interaction is known, we build on a single extended AR model to define time-domain and spectral GC measures which are non-negative and comprehensive of the whole  interdependencies existing in the bivariate process.


Our formulation assumes that instantaneous interactions are causal and oriented along the direction $Y_n \rightarrow X_n$, and makes use of the extended AR model (\ref{eq_AR_inst}), depicted in Fig. 1(c), to measure GC along both directions.
Since the description of $Y$ is unaltered with respect to the strictly causal model ((\ref{eq_AR_inst_b}) and (\ref{eq_VAR_b}) are identical), the time-domain and spectral GC measures along the direction $X\to Y$ are unchanged, corresponding to $F_{X\to Y}$ and $\tilde{f}_{X\rightarrow Y}(\Omega)$ defined in  
 (\ref{GCtimea}) and (\ref{fxy_ext}).

On the other hand, the time-domain GC along the direction $Y\to X$ is obtained combining the description of $X$ in (\ref{eq_AR_inst_a}) with its individual AR model in (\ref{eq_AR_restricted}a); elaborating the innovation variances yields the definition:
\begin{equation}\label{eq:extended_gc}
G_{Y\rightarrow X}
\triangleq
\ln\frac{\lambda_1^2}{\tilde{\sigma}_1^2}.
\end{equation}
Expanding (\ref{eq:extended_gc}) as $G_{Y\rightarrow X}=\ln\frac{\lambda_1^2}{\sigma_1^2}+\ln\frac{\sigma_1^2}{\tilde{\sigma}_1^2}$ and recalling (\ref{GCtimeb}) and (\ref{tildasigma1}, \ref{ICtime}) yields 
\begin{equation}\label{eq:extended_gcsum}
G_{Y\rightarrow X}= F_{Y\rightarrow X}+F_{Y\cdot X}
\end{equation}
which demonstrates that the use of the extended AR model (\ref{eq_AR_inst}) returns an \emph{extended} GC measure in the direction $Y\to X$ which incorporates both time-lagged and instantaneous causality, while preserving GC in the direction $X\to Y$.
We note also that the total coupling defined as the sum of the two GC measures in (\ref{eq:extended_gc}) is the same as that obtained by the strictly causal model (i.e., $F_{X\to Y} + G_{Y\to X} = F_{X;Y}$), showing that
orthogonalization preserves the total interaction between the two processes.

The frequency-domain GC along the direction $Y\to X$ is obtained following the treatment described in Sect. II.B (eqs. \ref{eq: matrix_form}-\ref{PSD_S}), which leads to express the individual PSD of the target process $X$ as 
\begin{equation}\label{PXext}
P_X(\Omega) = |H_{11}(\Omega)|^2 \tilde{\sigma}_1^2 + |\tilde H_{12}(\Omega)|^2 \sigma_2^2,
\end{equation}
from which the spectral GC is defined as the logarithmic ratio between the target PSD and its non-causal part:
\begin{equation}\label{eq:extended_spec_gc}
g_{Y\rightarrow X}(\Omega)
\triangleq
\ln\frac{P_X(\Omega)}
{\left|H_{11}(\Omega)\right|^2\tilde{\sigma}_1^2}.
\end{equation}
\noindent A comparison of (\ref{eq:extended_spec_gc}) and (\ref{eq_fyx}) reveals that the extended spectral Granger causality measure $g_{Y\rightarrow X}$ can be expressed in terms of $f_{Y\rightarrow X}$ as  $g_{Y\rightarrow X}(\Omega)=f_{Y\rightarrow X}(\Omega) + \ln (\sigma_1^2 /\tilde{\sigma}_1^2)$. Therefore, using the extended AR model to compute both spectral GC measures distributes instantaneous causality over all frequencies in the direction $Y \to X$, while preserving the Geweke measure in the direction $X \to Y$ .

Remarkably, both the spectral GC measures obtained with this representation, i.e. $\tilde{f}_{X\rightarrow Y}(\Omega)$ and $g_{Y\rightarrow X}(\Omega)$, are non-negative at each frequency $\Omega$ and are such that their full-frequency integration returns the time-domain measures $F_{X\rightarrow Y}$ and $G_{Y\rightarrow X}$.
Moreover, since the two measures capture the whole interaction between $X$ and $Y$, their sum is a reliable spectral measure of the total coupling, which is defined as
\begin{equation} \label{GspectTOT}
    g_{X;Y}(\Omega) \triangleq \ln\frac{P_X(\Omega)P_Y(\Omega)}
{|H_{11}(\Omega)\tilde{H}_{22}(\Omega)|^2\tilde{\sigma}_1^2\sigma_2^2}.
\end{equation}
It is important to note that the coupling measure $g_{X;Y}(\Omega)$ generally differs from $f_{X;Y}(\Omega)$ defined in (\ref{f_XY_freq}), although full-frequency integration returns the time-domain coupling $F_{X;Y}$ for both measures. The properties of these spectral measures of total interdependence are investigated in the simulation reported in the following. 

\section{Theoretical Example}

\subsection{Simulation setting and analysis}

To investigate the properties of the proposed measures and, in particular, their ability to explicitly account for instantaneous interactions, we consider a simulated bivariate stochastic process with controlled time-lagged and instantaneous dependencies. Specifically, two stationary processes $X$ and $Y$ are generated according to the bivariate AR model depicted in Fig. \ref{fig:simRes}a, defined by the following equations: 
\begin{equation}
\begin{aligned}
X_n &=
a_1X_{n-1}+a_2X_{n-2}
+cY_n+(2-c)Y_{n-1}
+W_{1,n},\\
Y_n &=
b_1Y_{n-1}+b_2Y_{n-2}
+dX_{n-1}
+W_{2,n},
\end{aligned}
\label{eq:simulation_model}
\end{equation}

\noindent where $W_{1,n}$ and $W_{2,n}$ are zero-mean, mutually uncorrelated white Gaussian innovations with unit variance (the covariance matrix of the innovations is the identity matrix).
The self-dynamics of the two processes are defined by the AR coefficients $a_1,a_2$ and $b_1,b_2$, which determine in the $Z$ domain the presence of complex-conjugate poles with radii and normalized frequencies set to $\rho_1=\rho_2=0.9$ and $f_1/f_s=0.4$, $f_2/f_s=0.1$. The stochastic oscillations generated in this way are transmitted between the two processes through the interactions from $Y$ to $X$, which are both instantaneous (depending on the variable parameter $c$) and time-lagged (depending on $2-c$), and the interactions from $X$ to $Y$, which are exclusively time-lagged (depending on the fixed parameter $d=0.2$).
The system is studied by varying $c$ in the range $[0, 2]$, so as to shift progressively the interaction $Y\rightarrow X$  from purely lagged ($c=0$) to purely instantaneous ($c=2$).

Starting from the ground truth model (\ref{eq:simulation_model}), which is expressed as an extended model in the form of (\ref{eq_AR_inst}), the analysis is performed first determining the theoretical values of the parameters of the corresponding strictly causal model (\ref{eq_AR}) and of the individual AR models (\ref{eq_AR_restricted}); these steps are performed respectively through inverse Cholesky decomposition, and through a procedure which recovers arbitrarily large time-lagged self-dependencies (here, up to a lag $q=20$) based on a discrete-time Lyapunov formulation of the Yule-Walker equations of the strictly causal bivariate AR model (details can be found respectively in \cite{faes2013framework} and \cite{sparacino2025quantifying}). Then, following the formulations in Sect. II, the GC measures $F_{X\to Y}$ and $F_{Y\to X}$ defined in (\ref{GCtime}), the instantaneous coupling $F_{X\cdot Y}$ in (\ref{ICtime}), the extended GC measure $G_{Y\to X}$ in (\ref{eq:extended_gc}), and the total coupling $F_{X;Y}$ in (\ref{eq:gc_total}) are computed in the time domain; moreover,  the strictly causal GC measures $f_{X\to Y}(\Omega)$ (\ref{eq_fxy}) and $f_{Y\to X}(\Omega)$ (\ref{eq_fyx}), the instantaneous interaction measure $f_{X\cdot Y}(\Omega)$ (\ref{eq_finst}), the Geweke GC measures $\tilde{f}_{X\to Y}(\Omega)$ (\ref{fxy_ext}) and $\hat{f}_{Y\to X}(\Omega)$ (\ref{fyx_ext}), the extended GC $g_{Y\to X}(\Omega)$ (\ref{eq:extended_spec_gc}), as well as the coupling measures $f_{X;Y}(\Omega)$ (\ref{f_XY_freq}) and $g_{X;Y}(\Omega)$ (\ref{GspectTOT}), are computed in the frequency domain.

\subsection{Results and discussion}

\begin{figure*} [t!]
\centering
\includegraphics[scale=1]{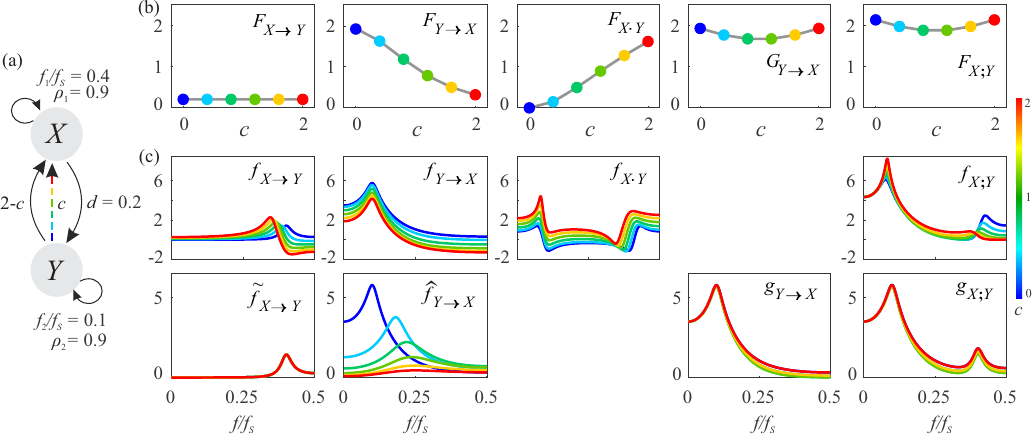}
\caption{Analysis of dynamic interactions in a simulated bivariate process with changing instantaneous and time-lagged effects.
(a) Schematic representation of the analyzed system, simulated to reproduce autonomous oscillations in $X$ and $Y$ (self-loops), time-lagged causal interactions from $X$ to $Y$ and from $Y$ to $X$ (solid arrows) and instantaneous effects oriented from $Y$ to $X$ (dashed arrow).
(b) Time-domain measures of GC from $X$ to $Y$ and from $Y$ to $X$ ($F_{X\rightarrow Y}$, $F_{Y\rightarrow X}$), instantaneous interaction ($F_{X\cdot Y}$), extended GC ($G_{Y\rightarrow X}$) and total interaction ($F_{X;Y}$), computed at varying the coupling parameter $c$ while fixing all other parameters (see text).
(c) Frequency-domain measures of GC ($f_{X\rightarrow Y}$, $f_{Y\rightarrow X}$, $\tilde{f}_{X\rightarrow Y}$, $\hat{f}_{Y\rightarrow X}$), instantaneous interaction ($f_{X\cdot Y}$), extended GC ($g_{Y\rightarrow X}$), and total coupling ($f_{X;Y}$, $g_{X;Y}$) derived from the strictly causal AR model (upper panels) and from the extended AR model (lower panels).
Panels of frequency-domain measures are aligned vertically with the corresponding time-domain measures, and colors indicate increasing values of $c$, from $c=0$ (blue) to $c=2$ (red).}
\label{fig:simRes}
\end{figure*}

The results of the analysis of the simulated system are depicted in Fig. \ref{fig:simRes}, reporting the time-domain measures estimated as a function of the coupling parameter (Fig. \ref{fig:simRes}b) and the corresponding frequency-domain measures (aligned vertically in Fig. \ref{fig:simRes}c).
The time domain measures follow faithfully the variations of the causal, instantaneous and total coupling expected at varying the parameter $c$. Specifically we see that, at increasing $c$, the GC $F_{X\to Y}$ is kept constant, while there is a decrease of the GC $F_{Y\to X}$ and a corresponding increase of the instantaneous interaction $F_{Y\cdot X}$; these trends reflect the unvaried interaction $X\to Y$ and the progressive shift from time-lagged to instantaneous of the interaction $Y\to X$. The opposite trends of time-lagged and instantaneous interaction in the direction $Y\to X$ determine small variations of the extended GC $G_{Y\to X}$, which are also reflected by the total coupling $F_{X;Y}$ due to the small and constant values of the GC $X\to Y$.

The frequency domain analysis confirms the theoretical properties of the various measures, and helps identifying their features and limitations. First, we verified analytically that all frequency-domain measures satisfy the spectral integration property, observing that full-frequency integration applied to each spectral measure (depicted by a colored line in Fig. \ref{fig:simRes}c) returns exactly the value of the corresponding time-domain measure (circle of the same color in Fig. \ref{fig:simRes}b). Second, as regards the detection of spectral GC, the analysis confirms that all approaches converge towards meaningful profiles in the absence of instantaneous effects: in this case ($c=0$, blue lines in Fig. \ref{fig:simRes}c), the spectral GCs $f_{X\to Y}=\tilde{f}_{X\to Y}$ and $f_{Y\to X}=\hat{f}_{Y\to X}=g_{Y\to X}$ are always non-negative and peak at 0.4 Hz and 0.1 Hz, reflecting the transfer of information from $X$ to $Y$ and from $Y$ to $X$ imposed at these frequencies. When the instantaneous interaction grows at the expense of the time-lagged effects ($c>0$), the spectral GC measures derived from the strictly causal model ($f_{X\to Y}$, $f_{Y\to X}$) keep informing about the time-lagged effects but display a shift in the peak frequency and negative values at some frequencies which make them more difficult to interpret. The measures derived from the Geweke method reflect accurately the time-lagged causality in the spectral domain: $\tilde{f}_{X\to Y}$ is unvaried, displaying the stable transfer of information from $X$ to $Y$ at 0.4 Hz, while $\hat{f}_{Y\to X}$ is smoothed progressively, reflecting the dampening of time lagged effects in this direction. On the other hand, the extended measure $g_{Y\to X}$ that incorporates both time-lagged and instantaneous transfer remains almost constant, reflecting in the spectral domain the opposite trends simulated for the time-domain measures at varying $c$.
Third, as regards the instantaneous and total coupling, the measure $f_{X\cdot Y}$ is of very difficult interpretation as it always takes negative values at some frequencies and does not exhibit clear positive peaks. Its main usefulness seems the fact that, summed to the two GC measures $f_{X\to Y}$ and $f_{Y\to X}$, returns the total coupling measure $f_{X;Y}$ that is never negative and peaks around the frequencies of the imposed oscillations (i.e. $f_1=0.4f_s$ and $f_2=0.1f_s$). A similar profile is provided by the alternative measure of total coupling $g_{X;Y}$, which is also non-negative and looks more stable in the sense that the peaks of the information transfer always occur at $f_1$ and $f_2$.

These results document that, in the time domain, the traditional strictly causal AR model captures all types of interactions (time-lagged via the two GC measures $F_{X\to Y}$ and $F_{Y\to X}$, instantaneous via $F_{X\cdot Y}$, and total via $F_{X;Y}$), and that the extended measure $G_{Y\to X}=F_{Y\to X}+F_{X\cdot Y}$ provides useful quantification of the overall (both time-lagged and instantaneous) causal interactions when the instantaneous effects are oriented along a known causal direction. On the other hand, the extended model is important to analyze interactions in the frequency domain when instantaneous effects are non-negligible. In this case, the measures derived from the strictly causal model ($f_{X\to Y}$, $f_{Y\to X}$ and $f_{X\cdot Y}$) have undesired properties (e.g., negativity) which impair their interpretation. Moreover, their sum returns a coupling measure which, though being closely connected to the coherence function (see (\ref{eq:coh})), may display peaks not well-aligned with the frequency of the underlying coupled oscillations. On the contrary, the extended model yields profiles of the spectral GC which are derived taking instantaneous effects in proper account. Specifically, the Geweke measures $\tilde{f}_{X\to Y}$ and $\hat{f}_{Y\to X}$ derived from different extended models return a meaningful spectral representation of purely time-lagged effects; while this is in the spirit of GC, the sum of the two measures lack of completeness and thus cannot be used as a spectral index of total coupling. Finally, the extended measure $g_{Y\to X}$ faithfully represents the presence and extent of both instantaneous and time-lagged effects in the spectral domain. When instantaneous dependencies can be assigned to a known causal direction, 
the use of a single extended model incorporating zero lag effects along this direction allows to reliably compute frequency-resolved measures of purely time-lagged causality and of extended (time-lagged and instantaneous) causality, respectively via $\tilde{f}_{X\to Y}$ and $g_{Y\to X}$, and to sum them to get a reliable measure of total coupling ($g_{X;Y}$). This is the approach followed in the practical applications presented in the next Section.

\section{Application to Cardiovascular Interactions}

\subsection{Experimental Data, Model Identification, and Statistical Analysis}

The study included 127 healthy young volunteers (75 females; age: $18.6 \pm 3.3$ years), all normotensive and with body mass index within the normal range ($21.4 \pm 2.2$ kg/m$^2$). Written informed consent was obtained from all participants before enrollment in the study, which was approved by the Ethics Committee of the Jessenius Faculty of Medicine, Comenius University, Martin, Slovakia.

\begin{figure} [t!]
\centering
\includegraphics[scale=0.9]{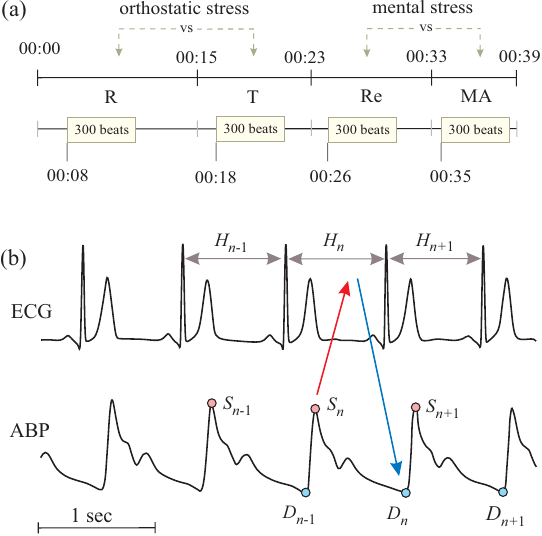}
\caption{(a) Timeline of the experimental protocol, indicating the sequence of the analyzed conditions - rest (R), head-up tilt (T), recovery after tilt (Re), and mental arithmetic (MA) - together with the timing of the 300-beat segments extracted from each condition for subsequent analysis and the two experimental comparisons considered in the study, the assessment of the response to orthostatic stress (R vs. T) and to mental stress (Re vs. MA). (b) Schematic representation of the beat-to-beat extraction procedure and indexing convention adopted for the cardiovascular variables. The $n$-th heart period, $H_n$, was defined as the interval between two consecutive R peaks of the electrocardiogram (ECG). The $n$-th systolic arterial pressure sample, $S_n$, was taken as the maximum value of the arterial blood pressure (ABP) signal measured within $H_n$, whereas the $n$-th diastolic arterial pressure sample, $D_n$, was taken as the minimum value of the ABP signal measured between the occurrence times of $S_n$ and $S_{n+1}$. The arrows indicate the instantaneous (within-beat) couplings implied by the adopted indexing convention, namely $S_n \rightarrow H_n$ (fast baroreflex effect) and $H_n \rightarrow D_n$ (fast run-off effect).}
\label{figure_intro}
\end{figure}

The experimental protocol was designed to study the autonomic activation related to postural stress and mental stress, and included the following consecutive phases: i) a 15-minute supine resting position (R); ii) a phase of head-up tilt (T), during which participants were tilted to $45^\circ$ for 8 minutes to induce mild orthostatic stress; iii) a supine recovery period (Re) lasting 10 minutes; and iv) a phase of mental arithmetic task (M) performed in the supine position for 6 minutes to induce mild mental stress. During the M task the subjects were instructed to sum up three-digit numbers (until one-digit number
was reached) displayed on the ceiling of the examination room by a data projector and to
decide if the final one-digit number was odd or even. After decision, the subject had to click
by computer mouse on the corresponding push button projected on the ceiling. During this phase subjects were disturbed by the rhythmic sound of a metronome and
instructed to perform the mental arithmetic task as quickly as possible with a minimal error
rate. The timeline of the study protocol is shown in Fig. \ref{figure_intro}a).

Throughout the protocol, the electrocardiogram (ECG, horizontal bipolar chest lead) and continuous arterial blood pressure (ABP, measured via the volume-clamp photoplethysmographic method) signals were acquired simultaneously at a sampling frequency of 1 kHz. From these recordings, the beat-to-beat time series of heart period (HP), systolic arterial pressure (SAP), and diastolic arterial pressure (DAP) were extracted as follows. The $n$-th HP value, denoted as $H_n$, was computed as the duration of the interval between the $n$-th and the $(n+1)$-th R peaks of the ECG detected using the Pan-Tompkins algorithm~\cite{pan1985real}. The corresponding $n$-th DAP and SAP values, denoted as $D_n$ and $S_n$, were computed as the minimum and maximum values, respectively, of the ABP waveform within the cardiac cycle associated with $H_n$, as illustrated in Fig.~\ref{figure_intro}b). With this approach, synchronous sequences of 300 consecutive points were obtained in each of the four experimental phases.
Prior to the analysis, the time series were high-pass filtered (IIR filter, cut-off frequency 0.0155 Hz) to remove slow trends and favor stationarity, and subsequently normalized to zero mean. A detailed description of the experimental protocol and signal acquisition procedure is available in~\cite{javorka2017basic, iovino2024comparison}.

The HP, SAP and DAP time series obtained for each subject and condition were regarded as realizations of the discrete-time stochastic processes $H=\{H_n\}$, $S=\{S_n\}$, and $D=\{D_n\}$, which were assumed as evenly sampled with a sampling frequency equal to the mean heart rate ($f_s=\frac{1}{\mathbb{E}[H_n]}$). Two separate analyses were performed, investigating the interaction between HP and SAP, and the interaction between HP and DAP; in each case, a bivariate AR model was used to describe the dynamic interactions between the two processes (respectively, $H$ and $S$, and $H$ and $D$).
In these two applications, instantaneous interactions are non-negligible, physiologically meaningful, and causally oriented: according to the adopted measurement convention (Fig.~\ref{figure_intro}b), the interaction between $H_n$ and $S_n$ is oriented along the direction $S_n\to H_n$, and represents fast (within-beat) baroreflex effects \cite{faes2014lag}, while the interaction between $H_n$ and $D_n$ is oriented along the direction $H_n\to D_n$, and represents the cardiac run-off \cite{javorka2017basic}. To properly represent these two cases by using the extended AR model as defined in (\ref{eq_AR_inst}), the analysis was performed by setting $X=H$ and $Y=S$ in the first case, and $X=D$ and $Y=H$ in the second case.

The analysis was performed by reducing the time series to zero mean and by following the procedure described in Sect. \ref{sec_methods}. For each pair of series, the parameters of the strictly causal AR model (\ref{eq_AR}) were estimated using the classical least-squares approach, while its order $p$ was selected according to the Akaike Information Criterion \cite{akaike1974new}. These parameters were then used to derive the parameters of the individual AR models (2) needed to compute the time-domain GC by formulating the Yule Walker equations and rearranging them to obtain autocorrelations up to a lag $q=20$ \cite{faes2016information, barnett2014mvgc}, and to derive the parameters of the extended bivariate AR model (\ref{eq_AR_inst}) by following the procedure described in Sect. II.B. The identified models were then used to estimate the five considered time-domain interaction measures, namely $F_{X\rightarrow Y}$, $F_{Y\rightarrow X}$, $F_{X\cdot Y}$, $F_{X;Y}$, and $G_{Y\rightarrow X}$. In the frequency domain, the measures capturing Granger causality based on the extended model (\ref{eq_AR_inst}), i.e. $\tilde{f}_{X\rightarrow Y}(\Omega)$ and $g_{Y\rightarrow X}(\Omega)$, as well as the measures of global coupling $f_{X;Y}$ and $g_{X;Y}$, were derived and integrated over the two main bands of physiological interest: the low-frequency band (LF, 0.03$-$0.15 Hz) and the high frequency band (HF, 0.15$-$0.40 Hz) \cite{cohen2002short}.

The statistical significance of all interaction measures estimated in both the time and frequency domains was assessed using surrogate data generated through the iteratively refined amplitude-adjusted Fourier Transform (iAAFT) procedure~\cite{schreiber1996improved}. The surrogate time series were constructed to retain the amplitude distribution and power spectrum of each process while eliminating any mutual coupling, according to the null hypothesis of uncorrelated processes with preserved individual dynamics. The statistical significance was evaluated using a non-parametric test based on percentiles, whereby each of the measures computed from the original time series was compared against its empirical distribution obtained from the surrogate realizations. A given measure was considered statistically significant when its original value exceeded the $95^{\text{th}}$ percentile of the surrogate distribution, while it was deemed not significant otherwise.

\subsection{Results and Discussion}

The results of the analysis are presented in Fig.~\ref{fig:HP_SAP} (HP$-$SAP interactions) and Fig.~\ref{fig:DAP_HP} (HP$-$DAP interactions), reporting the distributions of the interaction measures of coupling and causality computed across subjects both in the time domain and in the frequency domain (LF and HF bands) in the four analyzed conditions. As this study is focused on assessing the separately impact on cardiovascular interactions of postural stress and mental stress, the statistical analysis was performed by applying, for each considered interaction measure, the \textcolor{black}{non-parametric Wilcoxon signed-rank test (with significance levels set to $\alpha=0.05$ and $\alpha=0.001$)} on the distributions of the measure obtained during the rest and head-up tilt phases of the protocol (comparison R vs. T), and during the recovery and mental arithmetic phases (comparison Re vs. M).

\subsubsection{Heart Period - Systolic Pressure Interactions}

\begin{figure*} [t!]
\centering
\includegraphics[scale=0.95]{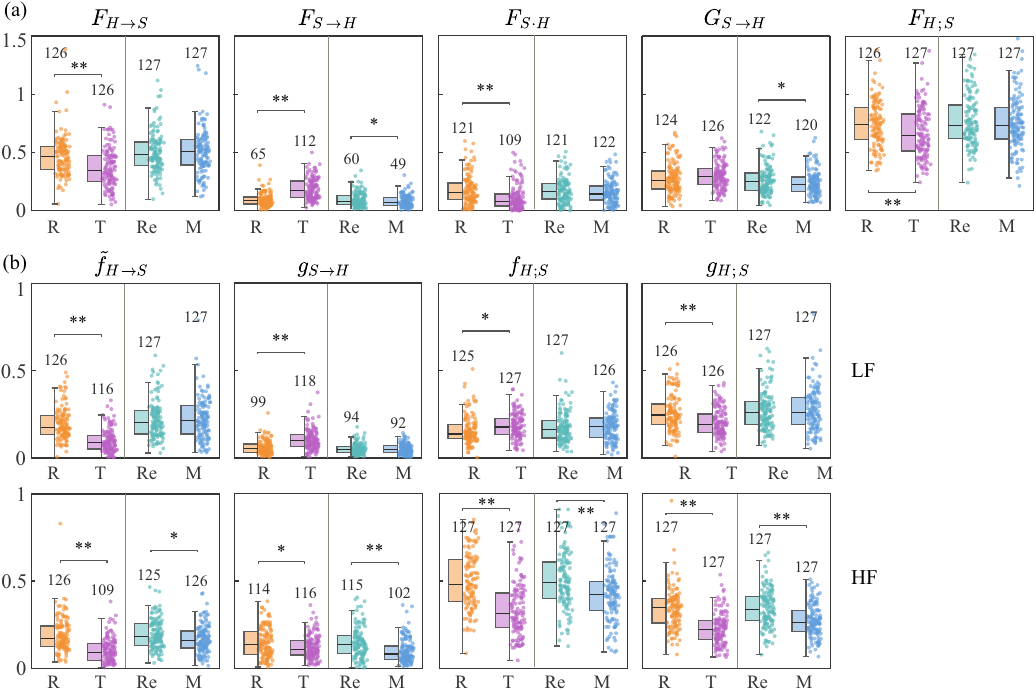}
\caption{Distributions of the measures characterizing the HP--SAP interaction in real cardiovascular data, shown as boxplots with individual subject values and the median indicated by the horizontal line within each box. Results are reported for two experimental comparisons, namely resting condition versus head-up tilt (R vs. T) and recovery versus mental arithmetic (Re vs. M). (a) Time-domain measures, including the GC from HP to SAP, $F_{H\rightarrow S}$, the GC from SAP to HP, $F_{S\rightarrow H}$, the instantaneous interaction $F_{S\cdot H}$, the extended GC from SAP to HP, $G_{S\rightarrow H}$, and the total interdependence $F_{H;S}$. (b) Corresponding frequency-domain measures, including $\tilde{f}_{H\rightarrow S}$, $g_{S\rightarrow H}$, and the total coupling measures $f_{H;S}$ and $g_{H;S}$, integrated over the low-frequency (LF) and high-frequency (HF) bands. The number of subjects showing statistically significant interactions according to the surrogate-based test is indicated above each boxplot, whereas asterisks indicate statistically significant differences between conditions ($^*p<0.05$, $^{**}p<0.001$, non-
parametric Wilcoxon signed-rank test).}
\label{fig:HP_SAP}
\end{figure*}

The time-domain analysis of the HP$-$SAP interaction during R and T (Fig.~\ref{fig:HP_SAP}a) shows that the total interdependence $F_{H;S}$ is statistically significant in almost all subjects, and decreases significantly from R to T. This trend is determined by a weakening of the causal interaction from HP to SAP (significant decrease of $F_{H\rightarrow S}$) observed together with unchanged interaction from SAP to HP (no significant changes of $G_{S\rightarrow H}$). In turn, the unvaried extended causality along the direction $S\to H$ is the result of markedly stronger time-lagged effects (marked increase of $F_{S\rightarrow H}$, both in magnitude and number of subjects in whom it was deemed as significant), observed with markedly weaker instantaneous effects (significant decrease of $F_{S\cdot H}$). These results confirm the tendency of postural stress to shift the closed-loop cardiovascular interactions towards the baroreflex feedback from SAP$\to$HP at the expense of the mechanical feedforward HP$\to$SAP, as observed in several previous works \cite{cooke1999human,faes2013mechanisms,javorka2017basic,javorka2018towards}. Moreover, the distinction between time-lagged and instantaneous effects allowed by our approach indicates that the well-known strengthening of baroreflex interactions induced by head-up tilt occurs exclusively for delayed effects (i.e., causal interactions $S\to H$ with lag $k\geq 1$), while fast baroreflex effects (i.e., within-beat causal interactions with lag $k=0$) are dampened significantly. \textcolor{black}{Physiologically, we interpret the decreased contribution of instantaneous effects with the higher heart rate observed for all subjects during tilt: with a shorter beat-to-beat interval, the time window allowed for baroreflex effects to occur becomes too short and the contribution of zero lag interaction in $S\to H$ direction is diminished.}

The corresponding frequency-domain analysis (Fig.~\ref{fig:HP_SAP}b) further clarifies these mechanisms, ascribing them to the oscillatory bands typical of cardiovascular variability. In particular, the tilt-induced baroreflex feedback activation (higher $g_{S\to H}$) and mechanical feedforward deactivation (smaller $\tilde{f}_{H\to S}$) are documented in the LF band of the spectrum, associated with sympathetic activity and the blood pressure Mayer waves \cite{cohen2002short}. \textcolor{black}{This result agrees with the expected dominant influence of the baroreflex on heart rate oscillations at the frequency of the Mayer waves.}
On the contrary, both spectral GC measures decrease significantly in the HF band, associated with parasympathetic and respiratory activities \cite{hedman1995high}. This pattern is more difficult to interpret as it likely reflects the indirect effects of respiration on both HP and SAP \cite{karemaker2009counterpoint}; nevertheless, one may speculate that lower causal and global effect in this band may be the result of the dampening of baroreflex-mediated respiratory sinus arrhythmia induced by tilt \cite{javorka2017basic,javorka2018towards}. \textcolor{black}{Taken together, the alterations in baroreflex interaction from $S$ to $H$ associated with tilt are dependent on the frequency of the oscillations (an increase in the LF band and a decrease in the HF band), highlighting the importance of assessing baroreflex function across different frequency bands and potentially improving our understanding of cardiovascular regulation.}

The analysis of the HP$-$SAP response during mental stress (Fig.~\ref{fig:HP_SAP}a) reveals substantially unchanged values of the total interdependence ($F_{H;S}$ does not vary significantly from Re to M), which are the result of a preserved feedforward interaction $F_{H\rightarrow S}$, which represents the dominant component of the HP$-$SAP coupling, preserved instantaneous interaction $F_{S\cdot H}$, and slight but statistically significant decrease of the time-lagged causality $F_{S\rightarrow  H}$ (which is also reflected in the extended measure $G_{S \rightarrow  H}$). These findings align with others in the literature \cite{javorka2017basic,javorka2018towards}, and indicate that the alterations of cardiovascular interactions induced by mental stress are more mild than those induced by postural stress, and of opposite nature: while tilt tends to increase the baroreflex coupling, mental arithmetic tends to decrease it. The decreased strength of the $SAP \rightarrow HP$ interaction during M could be related to an inhibition of the parasympathetic activity directed to the heart, since the baroreflex-mediated heart rate response is dominantly attributed to parasympathetic modifications. The corresponding frequency-domain analysis (Fig.~\ref{fig:HP_SAP}b) ascribes to the HF band the weaker coupling along the direction SAP$\to$HP observed moving from Re to M. 
\textcolor{black}{This reduction may reflect the attenuation of the baroreflex mechanism underlying respiratory sinus arrhythmia and $-$ more generally $-$ of vagally mediated cardiovascular oscillations typically associated with cognitive stress \cite{krohova2019multiscale}. Alternatively, the lower $SAP \rightarrow HP$ coupling observed at HF could be related to a shift in the breathing pattern toward more rapid and shallow breathing \cite{pitzalis1998effect} for which lower-magnitude respiration-induced blood pressure oscillations are expected.}


The measure of total dependence $g_{H;S}$ decreases significantly from R to T in both the LF and HF bands. As $g_{H;S}$ results exactly from the sum of its two causal components $\tilde{f}_{H\rightarrow S}$ and $g_{S\rightarrow H}$, the lower total coupling is explained by the decrease of both components in the HF band, and by a decrease of $\tilde{f}_{H\rightarrow S}$ which exceeds the increase of $g_{S\rightarrow H}$ in the LF band. In this band, an opposite trend is found for the measure $f_{H;S}$ related to the spectral coherence, which increases significantly from R to T. A similar tilt-induced increase of the LF coherence was observed previously \cite {nollo2005exploring}, marking a difference between the traditional methods to assess global coupling and the one proposed in this study.
Nevertheless, the two approaches yield similar results in the LF band regarding the transition from Re to M (unchanged total coupling), and in the HF band regarding both the transitions from R to T and from Re to M (significantly decreased total coupling). The latter result confirms that all types of cardiovascular interactions evaluated in the HF band decrease significantly their strength with both postural and cognitive challenges, suggesting a general stress-induced dampening of respiration-related and parasympathetic effects.


\begin{figure*} [t!]
\centering
\includegraphics[scale=0.95]{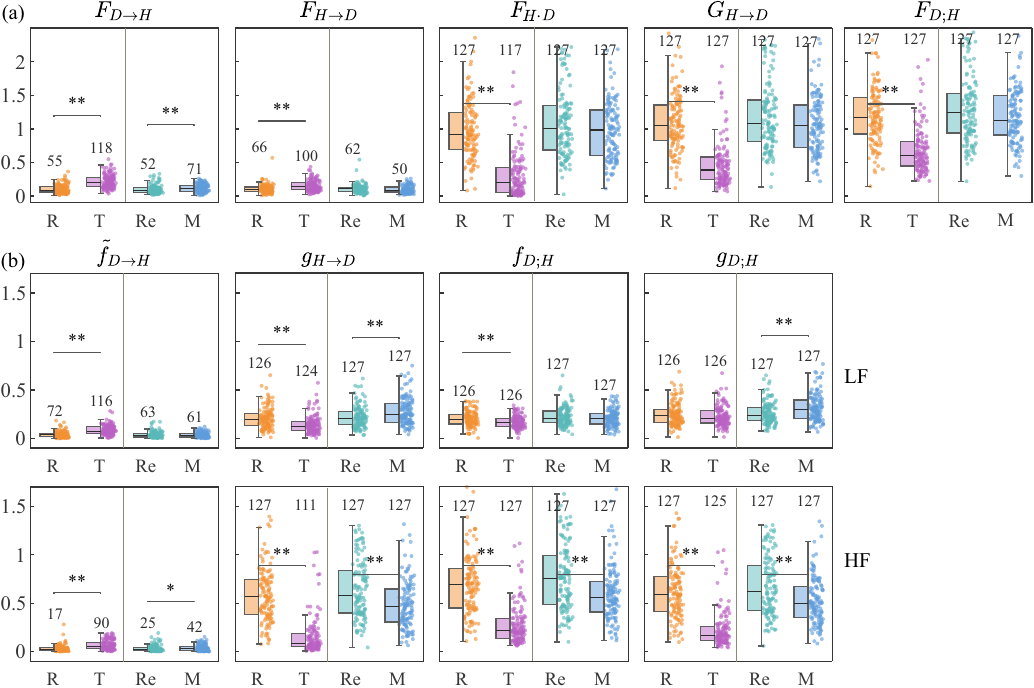}
\caption{Distributions of the measures characterizing the HP--DAP interaction in real cardiovascular data, shown as boxplots with individual subject values and the median indicated by the horizontal line within each box. Results are reported for two experimental comparisons, namely resting condition versus head-up tilt (R vs. T) and recovery versus mental arithmetic (Re vs. M). (a) Time-domain measures, including the GC from DAP to HP, $F_{D\rightarrow H}$, the GC from HP to DAP, $F_{H\rightarrow D}$, the instantaneous interaction $F_{H\cdot D}$, the extended GC from HP to DAP, $G_{H\rightarrow D}$, and the total interdependence $F_{D;H}$. (b) Corresponding frequency-domain measures, including $\tilde{f}_{D\rightarrow H}$, $g_{H\rightarrow D}$, and the total coupling measures $f_{D;H}$ and $g_{D;H}$, integrated over the low-frequency (LF) and high-frequency (HF) bands. The number of subjects showing statistically significant interactions according to the surrogate-based test is indicated above each boxplot, whereas asterisks indicate statistically significant differences between conditions ($^*p<0.05$, $^{**}p<0.001$, non-
parametric Wilcoxon signed-rank test).}
\label{fig:DAP_HP}
\end{figure*}

\subsubsection{Heart Period - Diastolic Pressure Interactions}

The time-domain analysis of the HP$-$DAP interactions (Fig.~\ref{fig:DAP_HP}a) shows that the interdependence between these two variables is dominated by the instantaneous effect $H_n \to D_n$, which reflects the known mechanism of cardiac run-off whereby diastole duration determines beat-by-beat diastolic pressure values \cite{baselli1994model,javorka2017basic}. Specifically, the instantaneous causality $F_{H\cdot D}$ is always statistically significant and prevails in magnitude over the two GCs $F_{D\rightarrow H}$ and $F_{H\rightarrow D}$ in all experimental conditions. The measure $F_{H\cdot D}$ is not altered by mental stress, while it markedly decreases during head-up tilt driving the corresponding significant drop of the total interdependence $F_{D;H}$. {Physiologically, a lower expression of the cardiac run-off during postural stress can be explained either by the rise of mechanisms able to influence DAP other than the cardiac run-off, or by the lower magnitude of HP oscillations occurring with tilt \cite{cooke1999human, javorka2017basic}.
On the other hand, the two GCs $F_{D\rightarrow H}$ and $F_{H\rightarrow D}$ are small and significant in only $\sim 50 \%$ of the subjects during R, Re and M, but increase significantly during T reaching statistical significance in $>80 \%$ of the subjects. This further confirms that the coupling between HP and DAP is dominated by the instantaneous effect, while the increase of $F_{D\rightarrow H}$ during postural and mental stress can be explained in terms of the activation of the baroreflex mechanism, tested in this case using DAP as the input signal \cite{cernanova2022input}; the stronger time-lagged interaction $H\to D$ agrees with a redistribution of the interaction lags (from dominance of lag-zero to emergence of time-lagged effects) previously observed in the transition from rest to tilt using information-theoretic non-uniform embedding methods \cite{javorka2017basic}.

The results of the frequency-domain analysis (Fig.~\ref{fig:DAP_HP}b) supports the time-domain findings. Considering the transition from R to T, a significant increase of $\tilde{f}_{D\rightarrow H}$ was observed together with a significant decrease of $g_{H\rightarrow D}$ in both the LF and HF bands. Methodologically, the increase of $\tilde{f}_{D\rightarrow H}$ follows that of $F_{D\rightarrow H}$, while the marked decrease of $g_{H\rightarrow D}$ reflects the dominant decrease of the instantaneous interaction which determines a downward shift of the whole profile of the extended GC (see Eqs. (13, 19, 31)).
As regards the transition from Re to M, the spectral analysis reveals that the overall stability of the DAP-HP interaction arises from opposite frequency-specific variations: the dominant interaction $g_{H\rightarrow D}$ increases significantly in the LF band and decreases significantly in the HF band, whereas the weaker interaction 
\textcolor{black}{$\tilde{f}_{D \rightarrow H}$ remains unchanged in the LF band and shows only a modest but significant decrease in the HF band, which can be attributed to a decreased parasympathetic activity-mediated baroreflex heart rate response. Here, since the dominant mechanical contribution of instantaneous effects is unchanged during mental stress, the frequency-specific variations of GC likely reflect the effect of alterations in the magnitude of the input signal (i.e., heart rate). More specifically, a change in the relative magnitude of heart rate oscillations in the assessed bands $-$ a relative increase in LF and a relative decrease in HF variability $-$ was observed in previous studies \cite{pagani1989spectral} potentially explaining the observed frequency-specific alterations in $g_{H \rightarrow D}$.}

The total coupling $g_{D;H}$ remains essentially unchanged from R to T in the LF band, as the increase in $\tilde{f}_{D\rightarrow H}$ is largely compensated by the decrease in $g_{H\rightarrow D}$. In the HF band, $\tilde{f}_{D\rightarrow H}$ increases while $g_{H\rightarrow D}$ decreases, with the larger magnitude of the latter determining the significant reduction of $g_{D;H}$. From Re to M, $g_{D;H}$ increases significantly in the LF band and decreases significantly in the HF band, following the frequency-specific changes of the dominant $g_{H\rightarrow D}$ component dominated by instantaneous cardiac run-off effects; these changes are also reflected by the coupling measure $f_{D;H}$.
The behavior of the latter measure differs significantly from that of $g_{D;H}$ only in the LF band, where a general tendency of $f_{D;H}$ to rise less in response to postural or mental stress is noticed.

\section{Concluding Remarks}
The main goal of this work was to establish precise guidelines for the assessment of coupling and causality measures in biomedical time series, considering either the entire dynamics or specific frequency bands of physiological interest. To this end, we systematically examined Granger causality and coherence measures within the linear signal processing framework \cite{granger1969investigating, geweke1982measurement,kay1988modern}, with particular emphasis on the role played by instantaneous dependencies, which are ubiquitous in physiological signals \cite{faes2013framework,porta2015wiener,pichot2024contribution}. Our results highlight important limitations of conventional measures and provide a consistent framework for addressing them when instantaneous dependencies have a physiologically meaningful direction.

For two processes $X$ and $Y$, we showed that causality and coupling can be properly quantified in the time domain using the standard strictly causal AR model commonly adopted to describe dynamic interactions. In fact, the contribution of instantaneous dependencies to the overall coupling is properly captured by $F_{X\cdot Y}$, which, together with the directional Granger causality measures $F_{X\to Y}$ and $F_{Y\to X}$, accounts for the total interdependence $F_{X;Y}$. Then, if the instantaneous dependency can be interpreted causally and its direction is known (e.g., $Y\to X$), $F_{X\cdot Y}$ becomes a measure of instantaneous causality and can be combined with $F_{Y\to X}$ to obtain the extended causality $G_{Y\to X}$ which accounts for both instantaneous and time-lagged effects.

The distinction becomes critical in the frequency domain. In the absence of instantaneous dependencies, the different formulations converge to the same spectral GC measures and satisfy the integration property linking the frequency- and time-domain representations. Nevertheless, when instantaneous dependencies are non-negligible the spectral GC measures based on the strictly causal AR model \cite{pernice2022spectral}, $f_{X\to Y}$ and $f_{Y\to X}$, may become negative at some frequencies, limiting their interpretation. 
On the other hand, the Geweke formulation \cite{geweke1982measurement}, being based on forcing instantaneous causality along the direction opposite to the investigated one, is appropriate to quantify the spectral GC $\tilde{f}_{X\to Y}$ when instantaneous causality has direction $Y\to X$; however, in this case the opposite Geweke spectral GC, $\hat{f}_{Y\to X}$, misses the instantaneous contribution.
We therefore propose using a single extended AR model to calculate spectral causality in both directions. The resulting measure $g_{Y\to X}$ accounts for both instantaneous and time-lagged causal effects, distributes the instantaneous contribution uniformly across frequencies, and integrates to the time-domain extended causality $G_{Y\to X}$. Thus, when the direction of instantaneous causality is physiologically meaningful and known \cite{javorka2017basic,faes2014lag}, this formulation provides a consistent representation of causal interactions across time and frequency domains. On the contrary, when instantaneous effects cannot be assigned to a physiologically meaningful causal directions (e.g., in the analysis of electroencephalographic data \cite{cao2022brain,koutlis2021comparison}), the Geweke formulation remains the most appropriate to represent time-lagged effects in the spectral domain.

As regards the detection symmetric dynamic interactions, the classical strictly causal AR formulation yields a spectral measure of total interdependence only through the definition of the spectral instantaneous interaction $f_{X\cdot Y}$ \cite{chicharro2011spectral,pernice2022spectral,faes2022new}. This quantity has limited interpretability and becomes negative at some frequencies even in the absence of instantaneous dependencies \cite{pernice2022spectral}. Its introduction in the frequency domain was motivated primarily by the desire to reproduce the time-domain decomposition \cite{chicharro2011spectral}, rather than by an independent mechanistic interpretation. We therefore see no compelling reason to retain a spectral instantaneous-interaction term when such an interpretation is unavailable, also because the spectral measure of total interdependence $f_{X;Y}$ can be equally quantified as defined in Eq. (17) by exploiting the well-known coherence function, which can be obtained directly from the spectral matrix without a parametric model \cite{kay1988modern,chicharro2011spectral}. 
As an alternative, we introduced $g_{X;Y}$, defined as the sum of the two spectral GC measures obtained from a single extended AR model. This measure retains the non-negativity and integration properties of $f_{X;Y}$ while showing a greater tendency to peak at frequencies corresponding to shared oscillatory components of $X$ and $Y$. It may therefore provide a useful complementary tool for identifying physiologically relevant frequency-specific coupling.

The application of the proposed framework to a typical context in the study of physiological signals, namely the analysis of cardiovascular interactions, further highlights the importance of accounting for instantaneous dependencies when characterizing dynamic coupling in both the time and frequency domains.
In the time domain, we showed that instantaneous dependencies play an important role in the analysis of SAP–HP interactions, where they reflect fast baroreflex effects \cite{faes2014lag}, and are even dominant in the analysis of HP–DAP interactions, where they reflect the cardiac run-off \cite{javorka2017basic}. Notably, because the strengths of time-lagged and instantaneous effects undergo opposite variations from rest to tilt, disregarding instantaneous dependencies would lead to opposite conclusions about the physiological response to postural stress of the baroreflex and cardiac run-off mechanisms.
In the frequency domain, properly accounting for instantaneous dependencies likewise provided a more refined characterization of the physiological mechanisms underlying low- and high-frequency cardiovascular oscillations. In the spectral analysis of HP–DAP interactions, we found that the dominant instantaneous interaction from DAP to HP, which is distributed across all frequencies, substantially shapes the corresponding patterns of spectral causality. In the spectral analysis of HP–SAP interactions, the newly proposed measure of total interaction revealed a different response to stress in the LF band compared with the measure based on classical coherence. This finding suggests an alternative interpretation of the balance between baroreflex and mechanical interactions in the presence of sympathetic activation.

Future studies should expand the methodological framework in several directions, including the extension of the proposed implementations to multivariate time series and the exploration of alternative approaches to the spectral analysis of causal interactions when instantaneous dependencies cannot be assigned a physiologically meaningful causal direction. Further physiological investigations are also warranted, particularly to clarify the role of respiration in the patterns of spectral causality underlying cardiovascular interactions in the high-frequency band.

\bibliographystyle{unsrt}
\bibliography{ref}

\end{document}